\documentclass[aps,prb,reprint,twocolumn,superscriptaddress,longbibliography]{revtex4-2}
\pdfoutput=1

\usepackage{graphicx}
\usepackage{amsmath,amsfonts,amssymb}
\usepackage{bbold}
\usepackage{physics}
\usepackage{color}
\usepackage{subfigure}
\usepackage[colorlinks=true,citecolor=blue,urlcolor=blue,linkcolor=blue]{hyperref}
\usepackage{array}
\usepackage{tabularx}
\usepackage{booktabs}
\usepackage{makecell}
\usepackage{textcomp}

\usepackage{bm}
\renewcommand{\Vec}[1]{\bm{#1}}

\begin{document}

\title{Fractional $1/3$ quantum vortices in chiral $d+id$ kagome superconductors}

\author{Frederik A. S. Philipsen}
\affiliation{Niels Bohr Institute, University of Copenhagen, DK-2200 Copenhagen, Denmark}

\author{Mats Barkman}
\affiliation{Niels Bohr Institute, University of Copenhagen, DK-2200 Copenhagen, Denmark}

\author{Andreas Kreisel}
\affiliation{Niels Bohr Institute, University of Copenhagen, DK-2200 Copenhagen, Denmark}
\affiliation{Department of Physics and Astronomy, Uppsala University, Box 516, 751 20 Uppsala, Sweden}

\author{Brian M. Andersen}
\affiliation{Niels Bohr Institute, University of Copenhagen, DK-2200 Copenhagen, Denmark}

\date{February 19, 2026}
\begin{abstract}
We perform a theoretical investigation of the nature of vortices in chiral $d+id$ superconductors on the kagome lattice. 
The study is motivated by recent experimental developments reporting evidence of time-reversal symmetry breaking in the superconducting state of kagome  metals. 
Using self-consistent microscopic calculations that incorporate the characteristics of the band structure of the kagome lattice, we find that fractional vortices permeate the ground state condensate in the presence of an external field. 
Each fractional vortex carries one third of the superconducting flux quantum and exhibits a characteristic signature related to one of the three sublattice degrees of freedom of the kagome lattice. 
We discuss the relevance of these results to recent experimental studies of kagome superconductors in the presence of an external magnetic field. 
\end{abstract}
\maketitle

\section{Introduction}

Superconductors which spontaneously break symmetries in addition to the standard global $U(1)$ gauge symmetry are typically described by a multicomponent order parameter. 
It is believed that these multicomponent superconducting states may be able to host topological excitations in the form of fractional vortices, carrying only a fraction of superconducting magnetic flux quantum $\Phi_0 = \frac{h}{2e}$ \cite{SigristUeda91}. 
These topological excitations play a key role in the superconductor's response to external fields and thermal fluctuations \cite{SigristPTP,babaev_2005, berg_2009}, leading to rich phases of matter.  

Fractional vortices have been a topic of great interest in the context of chiral triplet superfluidity as in superfluid ${}^3{\rm{He}}$ \cite{volovik_half_quantum_1985, salomaa_volovik_review_1987, volovik_book_2009, helium_3_half_quantum_experiment_2016,Makinen2019}, but also in the context of triplet superconductors \cite{Sauls01021994, tokuyasu_1990, Matsunaga, Ichioka2005,Chung_2009, babaev_garaud_2012_prl, babaev_garaud_2015, Haakansson2015,HasselbachUPt3}.
Even singlet superconducting states with spontaneous time-reversal symmetry breaking (TRSB) (e.g. $s+is$, $s+id$, $d+id$) are believed to be able to host fractional vortices \cite{solitons_garaud_carlstrom_babaev_2011,
CP2_garaud_carlstrom_babaev_2013,
Garaud2014,
CP2_benfenati_2023,
holmvall_2023_short, holmvall_2023_long}. In the chiral triplet $p+ip$ state, fractional vortices necessarily carry a half flux quantum (i.e. they are half-quantum vortices), enforced by the symmetry relating the two order parameter components to each other. 
A different kind of fractional vortices has been proposed to exist in multi-gap superconductors, associated with phase winding in only one of the gap components \cite{babaev_fractional_2002}.
Notably, in the absence of symmetries relating the two components, the magnetic flux associated with such fractional vortices may take any value between 0 and 1.
Such ``unquantized'' vortices were recently observed experimentally in the multiband superconductor ${\rm{Ba}}_{1-x}{\rm{K}}_x{\rm{Fe}}_2{\rm{As}}_2$ \cite{iguchi_2023, zhou_2024_BKFA_fractional_SQUID, zheng_2024_BKFA_fractional_STM}.

Even though multicomponent superconducting states can host fractional vortices, it does not necessarily mean that they are energetically  favorable compared to conventional Abrikosov vortices.
Superconducting states which spontaneously break time-reversal symmetry can take advantage of domain wall structures, reducing the energy cost of forming fractional vortices \cite{SigristPTP}. 
This makes studying the formation of fractional vortices particularly relevant in materials with TRSB superconductivity. Recent theoretical studies have investigated fractional vortices in $s+is$ superconductors from a self-consistent Bogoliubov-de Gennes (BdG) formalism, demonstrating the importance of maintaining microscopic degrees of freedom when predicting features of the fractional vortices \cite{CP2_benfenati_2023, iguchi_2023, Timoshuk_2024, Timoshuk_2025}. This includes the predicted magnetic flux associated with fractional vortex, which is not necessarily fully determined by the spontaneous symmetry-breaking of the ground state.

Recently, vanadium-based kagome metals $A$V$_3$Sb$_5$ ($A$: K, Rb, Cs) were discovered to be superconducting~\cite{Ortiz2020CsV3Sb5,OrtizEA21,YinEA21}. The kagome lattice is interesting since the distribution of sublattice weights of the eigenstates on the Fermi surface plays an important role in determining the leading instabilities arising from interactions~\cite{KieselEA12,Kiesel2013Unconventional,WuEA22,Scammell2023, Wu_2023_PDW_kagome,Ortiz2019New,Kenney2021Absence,Jiang2021Unconventional,Chen2021Roton,Zhao2021Cascade,ParkEA21,LinEA22,Denner2021,Tazai2022mechanism,Christensen2021,Ferrari2022,Christensen2022}. Regarding the preferred superconducting phase on the kagome lattice, several theoretical works have investigated Cooper pairing arising both from purely electronic fluctuations~\cite{Yu2012,KieselEA12,Wang2013,ParkEA21,Wu2021Nature,Tazai2022mechanism,Wen2022superconducting,Romer2022,Wu2021Nature,He2022Strong-coupling,LinEA22,Bai2022effective,Profe2024} and via phonon contributions~\cite{Wu2EA22,TanEA21,Zhang2021firstprinciples,Zhong2022Testing,Wang_phonon_2023,Ritz23}. From pairing via spin and charge fluctuations, the irreducible representation (irrep) with $d$-wave pairing symmetry stands out as a leading candidate~\cite{Romer2022}. Since this irrep is two-dimensional it will condense in a fully-gapped chiral  $d\pm id$ form. 

Experimentally, the nature of the superconducting ground state in the $A$V$_3$Sb$_5$ materials remains controversial~\cite{Wilsonreview}, with evidence for both standard nodeless non-sign-changing gaps and nodal unconventional superconducting order~\cite{Chen2021Roton,Liang2021Three-dimensional,Xu2021Multiband,Zhao2021nodal,Guguchia2022Tunable,Mielke2022Time-reversal}. Recent work has emphasized evidence for TRSB below the superconducting critical temperature $T_c$~\cite{Deng2024}, but it remains to be settled  whether the TRSB originates from intrinsic or extrinsic sources~\cite{Andersen2023,Li2021,Clara2022,Clara2023}. Ref.~\onlinecite{Deng2024} found quasi-particle interference patterns  dependent on the direction of the applied external magnetic field at $T<T_c$, consistent with the presence of a TRSB superconducting state. We stress that for the kagome lattice, standard phase-sensitiveness is wiped out by destructive sublattice interference effects. For example, as recently demonstrated theoretically, a chiral $d\pm id$ phase can exhibit features typically associated with conventional $s$-wave pairing, e.g. absence of in-gap impurity bound states, weak $T_c$-suppression rates with disorder, and even the presence of a Hebel-Slichter peak in the nuclear magnetic resonance (NMR) spin-lattice relaxation rate~\cite{Holbaek2023,YiDai2024, holbaek2026}. These highlighted properties are all relevant for the $A$V$_3$Sb$_5$ materials~\cite{RoppongiEA22,Xu2021Multiband,Zhang2023,Mu2021S-wave}. Thus, at present the chiral $d \pm id$ state remains a serious candidate for being realized in the $A$V$_3$Sb$_5$ materials, which has motivated part of our study. We stress, however, that the theoretical results presented in this paper are general and could apply to any kagome lattice featuring chiral TRSB superconductivity. 

\begin{figure}[t]
    \centering
\includegraphics[width=1\linewidth]{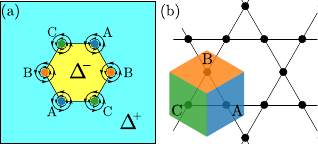}
    \caption{(a) Illustration of vortex state in a $d+id$ superconductor on the kagome lattice: Due to the interplay of the sublattice degrees of freedom (A,B,C), a phase boundary between the $d+id$ phase ($\Delta^+$) and a $d-id$ phase ($\Delta^-$) forms. At the boundary, a hexagonal structure with fractionalized vortices and suppressed order parameter on the three sublattices (blue, orange, green dots) form, exhibiting individual current loops around each of them. (b) Kagome lattice with the three sublattices A, B and C. The color and orientation of the diamonds are unique for each sublattice degree of freedom, and will be used throughout this work.
    }
    \label{fig:sketch}
\end{figure}

In this paper we investigate the nature of vortices in chiral $d+id$ superconductors on the kagome lattice. 
We employ a pairing mechanism which favors the $d+id$ state close to the van Hove singularities in the kagome bandstructure. Next, by using a microscopic and self-consistent BdG framework, we study the inhomogeneous superconducting state in the presence of an external magnetic field. 
We find states which fundamentally differ from  conventional Abrikosov vortices.
These states can be thought of as being composed by several fractional vortices, an example of which is illustrated in FIG.~\ref{fig:sketch}.
This state, which carries two superconducting flux quanta, is composed by a closed domain wall (separating an inner $\Delta^-=d-id$ phase from the outer $\Delta^+=d+id$ phase in the bulk), hosting six fractional vortices. 
Importantly, each fractional vortex can be associated with a specific sublattice degree of freedom of the kagome lattice, and carries each a third of the superconducting flux quantum, i.e. $h/6e$. 
Such 1/3 fractional vortices cannot be explained by a simple two-component superconducting order parameter, and illustrates the importance of the microscopic degrees of freedom and associated symmetries to fully understand the properties of the chiral superconducting state on the kagome lattice.

The paper is structured as follows. In Sec.~\ref{sec:methods} we present the methodology. More specifically, Sec.~\ref{subsec:TBH} introduces the kagome lattice, its tight-binding Hamiltonian, and the important characteristics of the sublattice degrees of freedom.
In Sec.~\ref{sec_sc} we introduce the interaction Hamiltonian and its mean-field decoupling which supports chiral $d$-wave superconducting order. Sections~\ref{sec_selfcon} and \ref{sec:peierls} contain details of the self-consistency procedure and the implementation of Peierls phases in the presence of an external magnetic field. Section \ref{sec:results} contains the main results of the paper.
We analyze various quantities, in support of the interpretation of these fractional vortices as belonging to individual sublattice degree of freedom.
We investigate how the solutions depend on the direction of the external magnetic field, and the difference between superconducting $d+id$ state close to the the upper and lower van Hove singularities. Finally, Sec. \ref{sec:conclusions} contains our discussion and conclusions.

\section{Methodology}\label{sec:methods}
\subsection{The electronic structure of the kagome lattice}\label{subsec:TBH}
\begin{figure}
    \centering
    \includegraphics[width=1\linewidth]{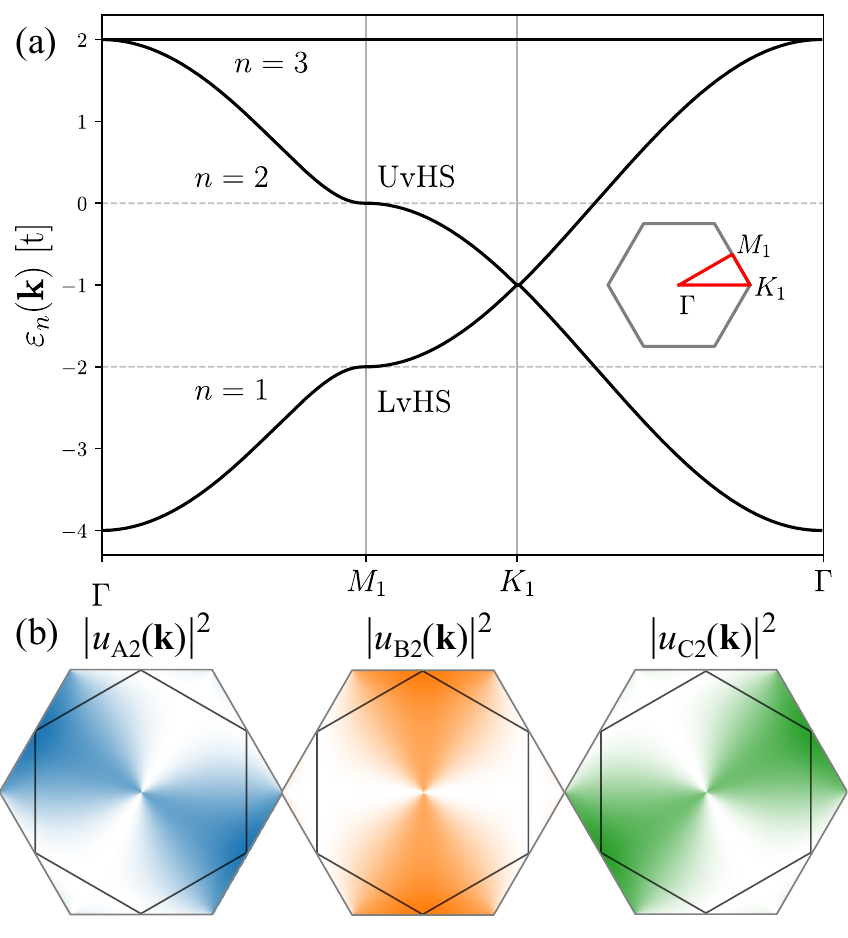}
    \caption{(a) Band structure of the kagome lattice along the high-symmetry path (see inset) for the three bands. The upper (lower) van Hove singularities are labeled as UvHS (LvHS). (b) Weight of the Bloch states in the second band ($n=2$) on the A, B and C sublattices displayed in the first Brillouin zone. The Fermi surface for $\mu=0$ is indicated in black.
    }
    \label{fig:Basic_kagome_plots}
\end{figure}

The kagome lattice structure consists of corner-connecting triangles formed by a triangular Bravais lattice with three sublattices per unit cell, see FIG.~\ref{fig:sketch}(b). Using a tight-binding model with only nearest-neighbor (NN) hoppings, the normal state Hamiltonian becomes
\begin{equation}\label{Eq:H_TB}
    \mathcal{H}_{\rm TB} =  
    -t \sum_{\substack{<\textbf{r}\textbf{r}'>},\sigma} 
    c_{\textbf{r}'\sigma}^{\dagger} c_{\textbf{r}\sigma}^{} -\mu \sum_{\textbf{r},\sigma} c_{\textbf{r}\sigma}^{\dagger} c_{\textbf{r}\sigma}^{}.
\end{equation}
Here $t$ is the nearest neighbor hopping, $\mu$ the chemical potential, and $\textbf{r}$ is a composite index $\textbf{r}=(\textbf{R},\alpha)$ where $\alpha \in \lbrace \mathrm{A},\mathrm{B},\mathrm{C} \rbrace$ denotes the sublattice degree of freedom and $\textbf{R}$ is the unit cell index, $\sigma$ denotes spin and $< \textbf{r}\textbf{r}'>$ nearest neighbor pairs. 
In reciprocal space, the Hamiltonian reads
\begin{equation}
     \mathcal{H}_{\rm TB} 
    = 
    \sum_{\sigma,\textbf{k} \in \rm{BZ}}
    \psi_{\textbf{k}\sigma}^{\dagger}
    H_{\rm TB}(\textbf{k})
    \psi_{\textbf{k}\sigma}^{},\label{Eq:Matrix_H_Tb}
\end{equation}
where $\psi_{\textbf{k}\sigma} = \begin{pmatrix} c_{\textbf{k}\mathrm{A}\sigma} & c_{\textbf{k}\mathrm{B}\sigma} & c_{\textbf{k}\mathrm{C}\sigma} \end{pmatrix}^T$
and the matrix $H_{\rm TB}(\textbf{k})$ is given by
\begin{equation}\label{Eq:H_momentum_normal_state}
    H_{\rm TB}(\textbf{k})
    =-
    \begin{pmatrix}
    \mu & 2t\cos(k_3) & 2t\cos(k_1)\\
    2t\cos(k_3) & \mu & 2t\cos(k_2) \\
    2t\cos(k_1) & 2t\cos(k_2) & \mu
    \end{pmatrix}.
\end{equation}
Here $k_i = \textbf{k}\cdot\textbf{a}_i$, with $\textbf{a}_1 =\frac{1}{2} \begin{pmatrix} 1 &0 \end{pmatrix}^T$, $\textbf{a}_2 =\frac{1}{2} \begin{pmatrix} \frac{1}{2} & \frac{\sqrt{3}}{2} \end{pmatrix}^T$ and $\textbf{a}_3 =\frac{1}{2} \begin{pmatrix} -\frac{1}{2} & \frac{\sqrt{3}}{2} \end{pmatrix}^T$ being the sublattice vectors. 

Diagonalizing the Hamiltonian using the unitary transformation 
$u_{\alpha n}^{*}(\textbf{k}) \hat{H}_{\mathrm{TB},\alpha\beta}(\textbf{k}) u_{\beta m}(\textbf{k}) = \varepsilon_n \delta_{nm}$, the three electronic bands $\varepsilon_n(\textbf{k})$ are obtained, as shown in FIG.~\ref{fig:Basic_kagome_plots}(a), along with the eigenstates of the $n$th band, $u_{\alpha n}(\textbf{k})$. The band structure of the kagome lattice features a flat band ($n=3$), a Dirac crossing at the $K$ point, and two van Hove singularities at the $M$ points in the $n=2$ and $n=1$ bands, respectively.
Furthermore, the $n=1$ and $n=2$ bands host a phenomenon known as sublattice interference, where the weight of the eigenstates $|u_{\alpha n}(\textbf{k})|^2$ on a given sublattice gains significant momentum dependence. FIG.~\ref{fig:Basic_kagome_plots}(b) displays the weights $|u_{\alpha n}(\textbf{k})|^2$ of the electronic states in the $n=2$ band on each sublattice, plotted in the first Brillouin zone. For both bands this phenomenon is especially strong near the van Hove singularities. 
When the Fermi surface is located precisely at the upper van Hove singularity (UvHS), the Bloch states are completely localized on only one sublattice at the $M$ points, while deviations from these points result in a non-zero weight on one additional sublattice.
The third sublattice weight remains zero until the next $M$ point is reached, meaning that at least one of the three sublattice weights is zero everywhere on the Fermi surface.
Exploring the vortex states at this filling will be the main focus of this work, however, the states formed in vicinity to the lower van Hove singularity (LvHS) will be explored in Sec.~\ref{sec:vortex_LvH} for comparison.

\subsection{The superconducting state}
\label{sec_sc}
We model superconductivity through a mean-field treatment of the interaction Hamiltonian
\begin{equation}
    \mathcal{H}_{\rm int} = \sum_{\textbf{r},\textbf{r}'} V^{}_{\textbf{r}\textbf{r}'} \;
    c_{\textbf{r}\uparrow}^{\dagger} c_{\textbf{r}'\downarrow}^{\dagger} 
    c_{\textbf{r}'\downarrow}^{} 
    c_{\textbf{r}\uparrow}^{},
\end{equation}
where $V^{}_{\textbf{r}\textbf{r}'} < 0$ is an attractive interaction potential. Mean-field decoupling in the Cooper channel, the
Hamiltonian can be rewritten in Nambu space as the Bogoliubov-de Gennes (BdG) Hamiltonian
\begin{equation}\label{Eq:H_BdG_real_space}
    \mathcal{H}_{\rm BdG} = 
    \Psi^{\dagger} 
    \begin{pmatrix}
        H_{\rm TB} & 
         \Delta \\
        \Delta^{\dagger} & -H_{\rm TB}^T\\
    \end{pmatrix}   
    \Psi,
\end{equation}
where $\Psi = \begin{pmatrix}
c_{(1,\mathrm{A})\uparrow}^{}& ...& c_{(N,\mathrm{C})\uparrow}^{}& c_{(1,\mathrm{A})\downarrow}^{\dagger} & ...& c_{(N,\mathrm{C})\downarrow}^{\dagger}
\end{pmatrix}^T$ denotes the real-space Nambu spinor for a system of $N$ unit cells. $H_{\rm TB}$ denotes the $3N\times3N$ matrix form of Eq.~\eqref{Eq:H_TB} and $\Delta$ is a $3N\times3N$ matrix with the entries given by self-consistency equation
\begin{equation}\label{Eq:Real_space_Delta_dif}
    \Delta_{\textbf{r}\textbf{r}'} = V_{\textbf{r}\textbf{r}'} \langle 
    c_{\textbf{r}'\downarrow} c_{\textbf{r}\uparrow}  \rangle,
    \qquad 
\end{equation}
where $\langle ...\rangle$ denotes the expectation value. 
Let us assume that the interaction potential $V_{\textbf{r}\textbf{r}'}$ is translation invariant, in the sense that it only depends on the relative distance $|\textbf{r} - \textbf{r}'|=\delta$.
For a given interaction distance $\delta$, we can classify the possible homogeneous superconducting states using the symmetry of the Kagome lattice, described by the $D_{6h}$ point group and its irreps. Such a decomposition was carried out in the case of on-site and nearest-neighbor interactions in Ref.~\cite{Holbaek2023}. The irreps associated with the lowest harmonics $s$, $p$, and $d$ are $A_{1g}$, $E_{1u}$, and $E_{2g}$, respectively, where $E_{1u}$ and $E_{2g}$ are two-dimensional irreps.
We want to consider a pairing interaction which through self-consistent calculations, favors $E_{2g}$ ($d$-wave).
In particular we are interested in stabilizing such states close to the van-Hove singularities.
This is relevant both from a theoretical perspective and for connection to experiments.
Several known superconducting kagome metals feature van-Hove singularities close to the Fermi level~\cite{Ortiz2020CsV3Sb5,OrtizEA21}, and it is precisely for these cases the sublattice interference is strong. We find that neither on-site or nearest neighbor pairing favors $E_{2g}$ close to the van-Hove singularities, and must therefore resort to longer-range interactions.

In this work, we focus on next-next nearest-neighbor (NNNN) pairing.
It is of particular interest, since it is the shortest range pairing (apart from on-site) which supports intra-sublattice pairing, i.e. pairing between the same sublattice degree of freedom. 
Close to the UvHS, where the sublattice interference is strong and the Bloch sublattice weights are extremely polarized (see FIG.~\ref{fig:Basic_kagome_plots}(b)), intra-sublattice pairing may be the dominant pairing mechanism~\cite{Romer2022}.
It turns out, as we will see later, that NNNN pairing favors chiral $d$-wave order close to the van-Hove singularities.

\begin{figure}
    \centering
    \includegraphics[width=1\linewidth]{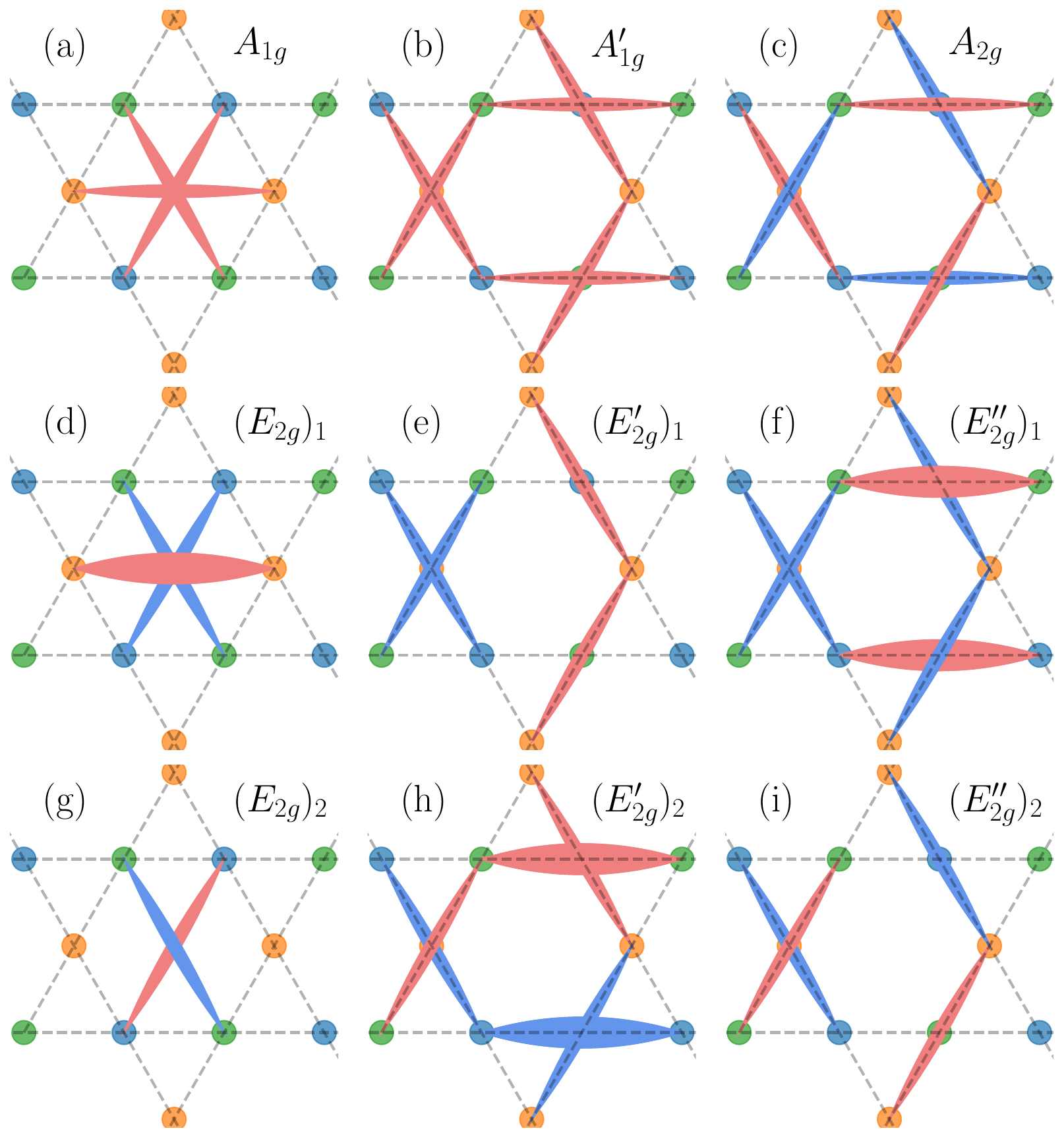}
    \caption{Visualization of the real-space representations (reps) describing the allowed singlet pairing states assuming only NNNN interactions. 
    The reps are denoted by the irreducible reps (irrep) which they transform as.
    We distinguish between reps transforming as the same irrep $\Gamma$ as $\Gamma'$ and $\Gamma''$. 
    For the two-dimensional reps ($E_{2g}$, $E_{2g}'$ and $E_{2g}''$) the outer index enumerates the components of the rep.
    The color of the line indicates the sign of $\Delta_{\textbf{r}\textbf{r}'}$: red positive and blue negative. The line thickness indicates the magnitude of $\Delta_{\textbf{r}\textbf{r}'}$: no line $0$, thin line $1$, and line thick $2$.
    }
    \label{fig:Visulisation_of_vector_rep}
\end{figure}

A brief symmetry classification, extending the analysis of Ref.~\cite{Holbaek2023} to NNNN pairing, will be presented below.
Each site has six next-next nearest neighbor bonds, for which we define superconducting bond parameters according to Eq.~\eqref{Eq:Real_space_Delta_dif}. 
For each sublattice, three of these bonds are related to the remaining three by translation, leaving us with nine independent bonds per unit cell, each having a singlet and triplet component. Further inspecting these bonds, we find two distinct types of bonds: the three inside-hexagon bonds are not related to the six out-of-hexagon bonds by any combination of translations or allowed point group operations. The inside-hexagon and out-of-hexagon bonds should be treated as two independent sets of bonds.

Decomposing this NNNN interaction in a similar way as in Ref.~\cite{Holbaek2023}, the singlet components of the inside-hexagon bonds decompose into $A_{1g}\otimes E_{2g}$, while the triplet components decompose into $ B_{1u}\otimes E_{1u}$. 
For the out-of-hexagon bonds, the singlet components decompose into $A_{1g}\otimes A_{2g}\otimes 2E_{2g}$ and the triplet into $B_{1u}\otimes B_{2u}\otimes 2E_{1u}$, where $2$ denotes two representations (reps) transforming as the same irrep.  
Reps of these irreps consist of linear combinations of the superconducting bond parameters, depicted in FIG.~\ref{fig:Visulisation_of_vector_rep} for the singlet bonds.
The color indicates the sign, red $+$ and blue $-$, while the line thickness indicates the magnitude factor; no line $0$, thin $1$ and thick $2$. 
This decomposition shows that $d$-wave superconducting order is possible from NNNN pairing, with the bond parameters on the inside-hexagon bonds being described by the $E_{2g}$ rep, and the order parameter on the out-of-hexagon being described by the two reps $E_{2g}'$ and $E_{2g}''$. 
\subsection{Self-consistency method and properties of the homogeneous phase}\label{sec_selfcon}
To determine the energetically favorable superconducting state we solve the self-consistency equation numerically.
For inhomogeneous solutions, we use the real space formulation and solve Eq.~\eqref{Eq:Real_space_Delta_dif}.
For homogeneous solutions, we can write the Hamiltonian in reciprocal space as
\begin{equation}\label{Eq:BdG_momentum_space}
    \mathcal{H}_{\rm BdG} = 
    \sum_{\textbf{k}} 
    \Psi_{\textbf{k}}^{\dagger}
    \begin{pmatrix}
        H_{\rm TB}(\textbf{k}) & 
         \Delta_{\textbf{k}} \\
        \Delta_{\textbf{k}}^{\dagger} & -H_{\rm TB}^T(-\textbf{k})\\
    \end{pmatrix}   
    \Psi_{\textbf{k}}^{},
\end{equation}
where $\Psi_{\textbf{k}}=\begin{pmatrix} \psi_{\textbf{k}\uparrow}^{} & \psi_{-\textbf{k}\downarrow}^{\dagger} \end{pmatrix}^T$ and $\Delta_\textbf{k}$ is a $3 \times 3$ gap matrix. 
The gap matrix $\Delta_\textbf{k}$ can be expressed in terms of the symmetry-allowed reps $\Gamma$ as
\begin{equation}\label{Eq:Self_consistency_momentum}
    \begin{aligned}
        \Delta_{\textbf{k},\alpha \beta} & = V  \sum_{\Gamma } 
     g_{\textbf{k}\alpha \beta}^\Gamma 
    \sum_{\substack{\textbf{k}' \alpha'\beta'}} g_{\textbf{k}'\alpha' \beta'}^\Gamma
    \langle c_{-\textbf{k}\beta'\downarrow} c_{\textbf{k}\alpha'\uparrow} \rangle \\
    & =
    \sum_{\Gamma} 
      g_{\textbf{k}\alpha \beta}^\Gamma C^\Gamma,
    \end{aligned}
\end{equation}
where $g_{\textbf{k}\alpha \beta}^\Gamma$ is the from factor of $\Gamma$.

The self-consistency equation is solved iteratively using a Bogoliubov transformation of the BdG Hamiltonian in Eq.~\eqref{Eq:BdG_momentum_space} to calculate the expectation values $\langle c_{-\textbf{k}\beta'\downarrow} c_{\textbf{k}\alpha'\uparrow} \rangle$ and the updated gap matrix $\Delta_\textbf{k}^{(t+1)}$. Performing this procedure iteratively until an appropriate convergence condition is reached, here we use $|\Delta_{\textbf{k}}^{(t+1)}-\Delta_{\textbf{k}}^{(t)}|/|\Delta_{\textbf{k}}^{(t)}| \le 10^{-8}$. 
The gap matrix necessarily corresponds to an extrema in the free-energy landscape.
Verification of this extrema as a minimum, and further identification of the global minimum, is achieved by repeating the calculation with different initial conditions.

We find that for an attractive NNNN interaction, a chiral $d$-wave state is favored when the chemical potential is in the vicinity of both the upper and lower van-Hove singularities. 
These chiral $d$-wave states are linear combinations of all three reps transforming as the $E_{2g}$ irrep shown in FIG.~\ref{fig:Visulisation_of_vector_rep}.
The phase difference between the two component in each rep is $\pm \pi / 2$. 
To be explicit, two degenerate chiral states $\Delta_\textbf{k}^\pm$ can be written as
\begin{equation}\label{Eq:chiral_UvH_state}
    \begin{aligned}
    \Delta^{\pm}_{\textbf{k}} & =  
    C^{E_{2g}} 
    \Big(  
    g^{(E_{2g})_1}_{\textbf{k}} \pm 
    i g^{(E_{2g})_2}_{\textbf{k}} 
    \Big)
    \\
    &+
    C^{E_{2g}'}
    \Big(
    g^{(E_{2g}')_1}_{\textbf{k}} 
    \mp
    i g^{(E_{2g}')_2}_{\textbf{k}} 
    \Big)
    \\
    &+
    C^{E_{2g}''}
    \Big(
    g^{(E_{2g}'')_1}_{\textbf{k}} \pm 
    i g^{(E_{2g}'')_2}_{\textbf{k}} 
    \Big),
    \end{aligned}
\end{equation}
where $C^\Gamma$ are real scalar factors. For concreteness, near the UvHS using the parameters $\mu = -0.02t$, $k_B T = 0.02t$, and $V = -0.75t$, with a grid size of $200 \times 200$ $\textbf{k}$-points, we get $C^{E_{2g}} \approx 0.0322t$, $C^{E_{2g}'} \approx -0.0356t$ and $C^{E_{2g}''} \approx 0.0326t$. 
In the vicinity of the LvHS, with the parameters $\mu = -1.98t$, $k_B T = 0.02t$, $V = -0.95t$ and a similar grid size, the self-consistent solution has $C^{E_{2g}} \approx 0.0197t$, $C^{E_{2g}'} \approx0.0462t$, and $C^{E_{2g}''} \approx 0.0286t$.
When we later perform calculations at the UvHS or the LvHS, it is specifically these sets of parameter we refer to.
For calculations in real space, a smaller grid-size than $200 \times 200$ is used, introducing some finite-size effects that slightly modify the bulk values of $C$ for the individual $E_{2g}$ reps. 

The doubly degenerate chiral $d$-wave states at the UvHS and LvHS will be the central focus in the remainder of this work. Since the Fermi surfaces and the DOS at these two fillings are identical, any difference in the superconducting state directly reflects the distinct sublattice structure of the electronic states at the two van-Hove singularities.

\subsection{Magnetic field and Peierls phases}\label{sec:peierls}
To study the formation of vortices within the chiral superconducting phase, we use a real-space BdG formalism where the interplay with the orbital magnetic field is included through Peierls substitution \cite{Peierl}. That is, we let

\begin{equation}
    t_{\textbf{r}' \textbf{r}}  \mapsto t_{\textbf{r}' \textbf{r}} e^{i \varphi_{\textbf{r}' \textbf{r}}} ,
\end{equation}
where $\varphi_{\textbf{r}'\textbf{r}}$ is defined as
\begin{equation}\label{Eq:Peierls_phase_def}
    \varphi_{\textbf{r}'\textbf{r}} =
    -\frac{\pi}{\Phi_0} \int_{\textbf{r}}^{\textbf{r}'} \textbf{A}(\Vec{\ell}) \cdot  \dd{\Vec{\ell}},
\end{equation}
where $\textbf{A}$ is the magnetic vector potential and $\Phi_0 = h / 2e$ the superconducting flux quantum ($h$ is Planck's constant and $e > 0$ is the elementary charge). This coupling to the vector potential makes the Hamiltonian invariant under the gauge transformation
\begin{equation}
    \begin{aligned}
        c_\textbf{r} & \mapsto e^{i \phi_\textbf{r}} c_\textbf{r}, \\
        \Delta_{\textbf{r} \textbf{r}'} & \mapsto e^{i(\phi_\textbf{r} + \phi_{\textbf{r}'})} \Delta_{\textbf{r} \textbf{r}'}, \\
        \varphi_{\textbf{r}' \textbf{r}} & \mapsto \varphi_{\textbf{r}' \textbf{r}} + (\phi_{\textbf{r}'} -\phi_\textbf{r}), 
\end{aligned}
\end{equation}
or equivalently $\textbf{A} \mapsto \textbf{A} - \frac{\hbar}{e} \grad{\phi}$.
We will consider the case of a homogeneous magnetic field, which is a good approximation when the magnetic penetration length is much larger than the coherence length. This allows us to choose a gauge and calculate the Peierls phase once and only solve the superconducting link variables $\Delta_{\textbf{r} \textbf{r}'}$ self-consistently.

\begin{figure*}
    \centering
    \includegraphics[width = 1 \linewidth]{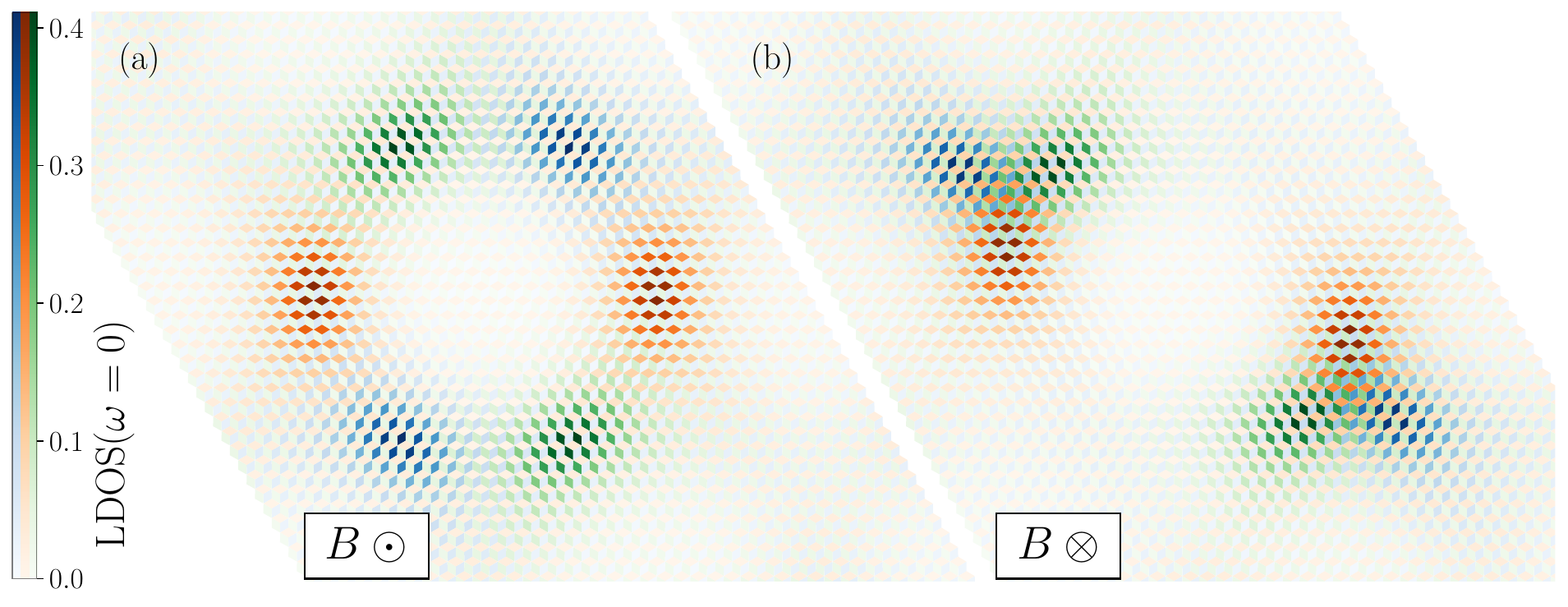}
    \caption{LDOS
    at the Fermi level for the vortex states found at the upper van Hove singularity.
    The magnetic field in (a) is in the out-of-plane ($\odot$ direction), and in (b) it is in the into-plane ($\otimes$ direction). 
    In both cases the bulk chirality is $\Delta^+$ and the magnetic flux through the system is $\pm 2 \Phi_0$. The orientation of each diamond along with the three different colormaps tells which sublattice degree of freedom each site belongs to (see Fig \ref{fig:sketch}(b) for reference).
    Although the structure is different for the two field directions, we clearly observe six regions of enhanced LDOS indicating the presence of six vortices. Each vortex carries a magnetic flux $\Phi_0 / 3$ and is associated with only one sublattice degree of freedom (A in blue, B in orange, C in green).
    }
    \label{fig:LDOS_UvH}
\end{figure*}

In order to capture the physics of the bulk superconducting state, we employ periodic boundary conditions instead of open boundary conditions. 
Periodicity fixes the allowed strengths of the magnetic field - the total magnetic flux through our magnetic unit cell must equal a multiple of the \textit{magnetic} flux quanta $h / e$ (i.e. two superconducting flux quanta), which can be derived formally using magnetic translation operators \cite{Zak_MTG_1964,Zak_MTG_2_1964,goto_magnetic_1996}.
This allows us to easily calculate the Peierls phases for each hopping term inside the magnetic unit cell. 
Special attention however, needs to be given to the hopping terms on the boundary corresponding to hopping into an adjacent magnetic unit cell.
These bonds acquire an additional phase factor such that the periodic boundary conditions mimic the translation of the magnetic unit cell.
Let $\Vec{\tau}$ be one of the translation vectors of the magnetic unit cell. Then the hopping term, which originally corresponded to hopping into the magnetic unit cell at $\Vec{\tau}$, should acquire the additional phase factor
\begin{equation} \label{eq: extra magnetic boundary phase}
    \delta \varphi_{\textbf{r}' \textbf{r}} =  \frac{\pi}{\Phi_0} \textbf{A}(\Vec{\tau}) \cdot \textbf{r}.
\end{equation}
This ensures that the magnetic flux through each boundary plaquette takes the same value as in the bulk of the magnetic unit cell.
\section{Results}\label{sec:results}

Motivated by experiments proposing vortices in a TRSB superconducting state~\cite{Deng2024}, we study the vortex states in chiral $d\pm id$ superconductivity on the kagome lattice.
Our main focus will be on the UvHS, complemented by studies at the LvHS in order to investigate how the characteristic sublattice interference influences the vortex states.

Two superconducting flux quanta are set to penetrate through a system of $40\times40$ unit cells, in which the superconducting order is initialized in the homogeneous $\Delta^+$ state.
The chiral superconducting state spontaneously breaks time-reversal symmetry such that positive and negative fields perpendicular to the plane are not degenerate, and both cases need to be considered separately.
We fix the superconducting bulk to be $\Delta^+$ and consider the two field directions. An alternative approach would be to fix the field direction and consider the two different bulk chiralities. Both approaches are equivalent and generate the same vortex states.

\subsection{Vortex states at the upper van Hove singularity}

\subsubsection{Local Density of States}

We start by presenting the two distinct states at the UvHS obtained from BdG calculations with the magnetic field in the out-of-plane $\mathbf B=|B|\mathbf e_z$ ($\odot$) and into-plane $\mathbf B=-|B|\mathbf e_z$ ($\otimes$) directions. For these systems, the local density of states (LDOS) is calculated, on the site $\textbf{r}$ at an energy $\omega$, using
\begin{equation}\label{Eq:Cal_LDOS}
    \mathrm{LDOS}(\textbf{r},\omega) = \frac{-1}{\pi} \mathrm{Im} \bigg\{ 
    \mathrm{Tr}\Big[ \mathcal{G}^{R}(\textbf{r},\omega) \Big]
    \bigg\},
\end{equation}
where $\mathcal{G}^{R}(\textbf{r},\omega)$ is the retarded Green's function
\begin{equation}
    \mathcal{G}^{R}(\textbf{r},\omega) = \big((\omega+i\eta) \: \mathbb{1} -H_{\rm{BdG}}\big)^{-1},
\end{equation}
where $H_{\rm{BdG}}$ is the matrix in Eq.~\eqref{Eq:H_BdG_real_space} and $\eta$ is a real valued infinitesimal, set to $\eta=0.035$ in this work.
The LDOS signature at the Fermi level are shown in FIG.~\ref{fig:LDOS_UvH}.
Each site is represented by a diamond whose orientation tells which sublattice degree of freedom it belongs to, see
FIG.~\ref{fig:sketch}(b). This allows us to easily extract sublattice-resolved information from the LDOS.

For the out-of-plane field ($\odot$) the LDOS displays a hexagonal shape which has six distinct regions of increased LDOS, see FIG.~\ref{fig:LDOS_UvH}(a).
From the use of the sublattice-specific colormaps, it is clear that each region of increased LDOS is predominantly on one of the three sublattices. Vortex cores come with regions of increased LDOS, which conventionally originates from the formation of spatially localized in-gap Caroli-de Gennes-Matricon states \cite{CAROLI_1964}. This suggest that the state shown in FIG.~\ref{fig:LDOS_UvH} has six separate vortices. 
Having restricted the magnetic flux through the system to two superconducting flux quanta, each of these six vortices is a fractional vortex and carries $1/3$ of a superconducting flux quantum. 

\begin{figure}[b]
    \centering
    \includegraphics[width = 1 \linewidth]{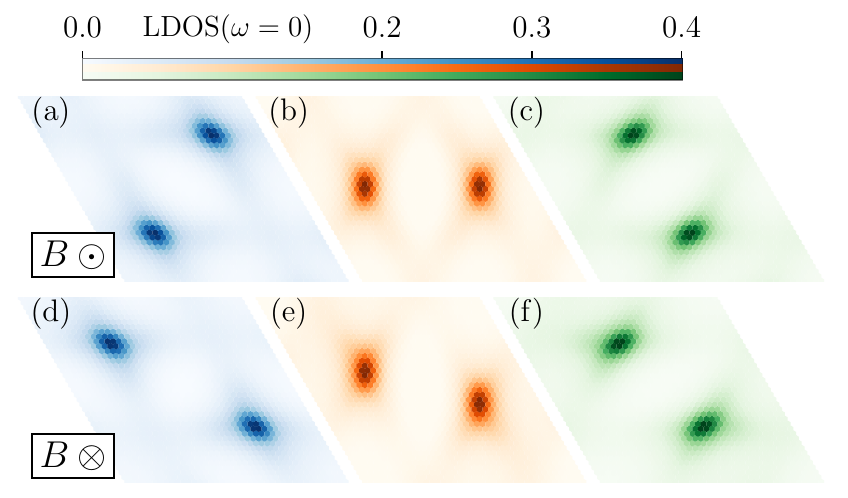}
    \caption{LDOS at the Fermi level in FIG.~\ref{fig:LDOS_UvH} shown separately for each of the sublattices. The top (bottom) row correspond to magnetic field in the $\odot$ ($\otimes$) direction, while each column (left to right) correspond to the A, B and C sublattices.
    }
    \label{fig:Sub_res_LDOS_UvH}
\end{figure}
\begin{figure*}
    \centering
    \includegraphics[width = 1 \linewidth]{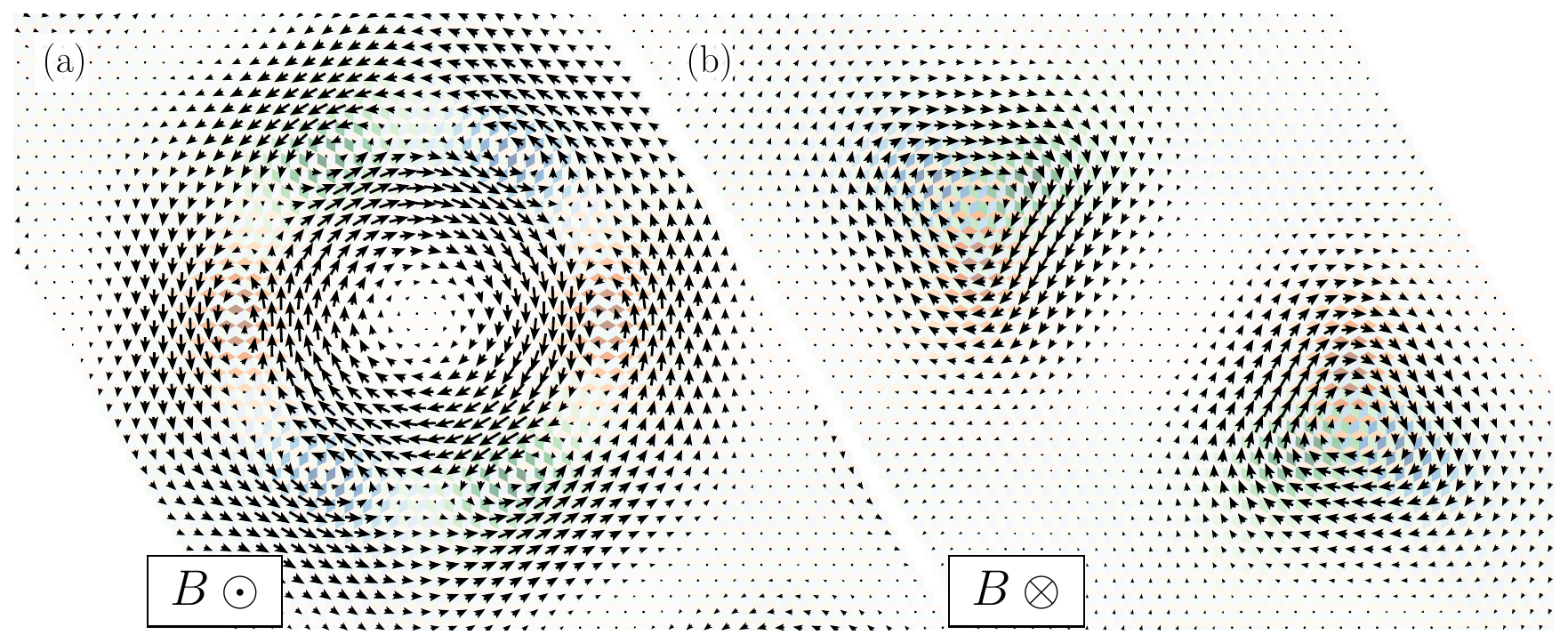}
    \caption{The vector fields display the current flowing through each unit cell in the vortex states at the UvHS for the $\odot$ (a) and $\otimes$ field directions. A faded version of the LDOS, shown in FIG.~\ref{fig:LDOS_UvH}, is plotted underneath to compare positions of LDOS enhancement with circulation of current.
    }
    \label{fig:J_unit_cell}
\end{figure*}
For the opposite field direction ($\otimes$), shown in FIG.~\ref{fig:LDOS_UvH}(b), a different LDOS is seen, where instead of one hexagon there are two triangular features. 
The LDOS is predominantly enhanced on a single sublattice near each corner of the triangles. 
This behavior suggests that each triangular feature does not correspond to a single conventional Abrikosov vortex, but instead consists of three tightly bound fractional vortices.
Similar to the previous field direction, these fractional vortices are predominantly associated with one of the three sublattice degrees of freedom and carry $-1/3$ of the superconducting flux quantum. The difference in the collective structure formed under the opposite field directions is expected from the fact that the bulk superconducting state is chiral and breaks time-reversal symmetry.

To further visualize the sublattice-specific nature of these vortices, the same LDOS is shown in FIG.~\ref{fig:Sub_res_LDOS_UvH} for the three sublattices separately. 
It is clear that the peaks in LDOS for each sublattice do not overlap, and where there is an enhanced LDOS for one sublattice, no enhancement of LDOS in the other two is observed.
Even though the different sublattices are connected through nearest neighbor hopping terms, the sublattice interference effectively decouples the sublattices and applies even on a local scale near the vortices.

\subsubsection{Current patterns}

Next we analyze the current patterns of the vortex states discussed in the previous paragraph. The current from site $\textbf{r}$ to $\textbf{r}'$ is given by
\begin{equation}\label{Eq:currents}
    J_{\textbf{r}' \textbf{r}} = 
    \frac{e}{\hbar} \mathop{\mathrm{Im}}
    \left(
    t_{\textbf{r}' \textbf{r}}
    e^{i \varphi_{\textbf{r}' \textbf{r}}}
    \langle c^\dagger_{\textbf{r}'\sigma} c_{\textbf{r}\sigma}^{} \rangle
    \right),
\end{equation}
and we define the current through site $\textbf{r}$ as
\begin{equation}\label{Eq:currents_pr_site}
    \Vec{J}_\textbf{r} = \sum_{\textbf{r}'} J_{\textbf{r}' \textbf{r}} \: \Vec{e}_{\textbf{r}' \textbf{r}},
\end{equation}
where $\Vec{e}_{\textbf{r}' \textbf{r}}$ is the unit vector pointing from $\textbf{r}$ to $\textbf{r}'$. 
The total current through a unit cell shown in FIG.~\ref{fig:J_unit_cell}, is calculated by summing $\Vec{J}_\textbf{r}$ over each site in the unit cell.
The current is displayed together with the LDOS from FIG.~\ref{fig:LDOS_UvH} in order to directly compare locations of currents and LDOS enhancements. 

The current in the out-of plane direction ($\odot$) is shown in FIG.~\ref{fig:J_unit_cell}(a) with pronounced clockwise (anti-clockwise) circulating currents on the interior (exterior) of the hexagonal structure formed by the six LDOS peaks.
A closer inspection reveals additional but crucial currents that circulate precisely the regions of increased LDOS, consistent with the interpretation of six fractional vortices.
Even though these calculations were performed assuming a homogeneous magnetic field, it is informative to discuss the inhomogeneous magnetic field induced by these currents.
The clockwise current on the interior of the hexagon will induce a magnetic field that opposes the external field.
The total magnetic field will be suppressed on the interior and large on the hexagonal boundary, with local maxima where the fractional vortices are located.

\begin{figure}
    \centering
    \includegraphics[width = 1 \linewidth,trim=0.4cm 0cm 1.2cm 0cm,clip]{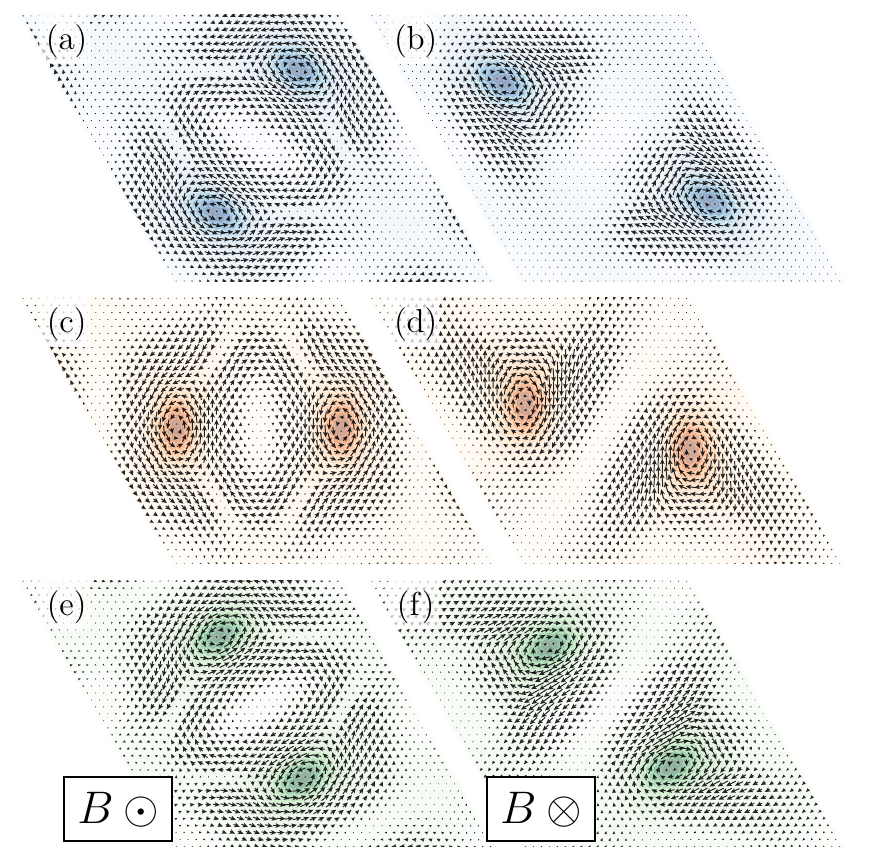}
    \caption{The vector field shows the current flowing over all sites of a specific sublattice for the vortex states found at the UvHS, with the LDOS on the corresponding sublattice underneath. 
    Here the rows (top to bottom) correspond to the three sublattices A, B and C, while the columns (left to right) refer to the two field directions out-of-plane $\odot$ and into-plane $\otimes$.
    For both field directions, we see that the sublattice-resolved current circulates around the point where there is an LDOS enhancement in the same sublattice.
    }
    \label{fig:J_sub}
\end{figure}

For the opposite field direction ($\otimes$), the current shown in FIG.~\ref{fig:J_unit_cell}(b) mainly circulates the center of each triangular feature.
At first glance, this gives no indication of the presence of the three individual fractional vortices in each, but is instead reminiscent of the currents associated with an Abrikosov vortex.
To investigate this further, the sublattice-resolved current is shown in FIG.~\ref{fig:J_sub}.
We see that the current through a particular sublattice circulates precisely around the location where an enhanced LDOS is seen on the same sublattice, in agreement with the fractional vortex interpretation.
Even though this sublattice resolved current cannot be observed experimentally, it is helpful in understanding the sublattice specific structure of the fractional vortices.

\begin{figure}
    \centering
    \includegraphics[width = 1 \linewidth]{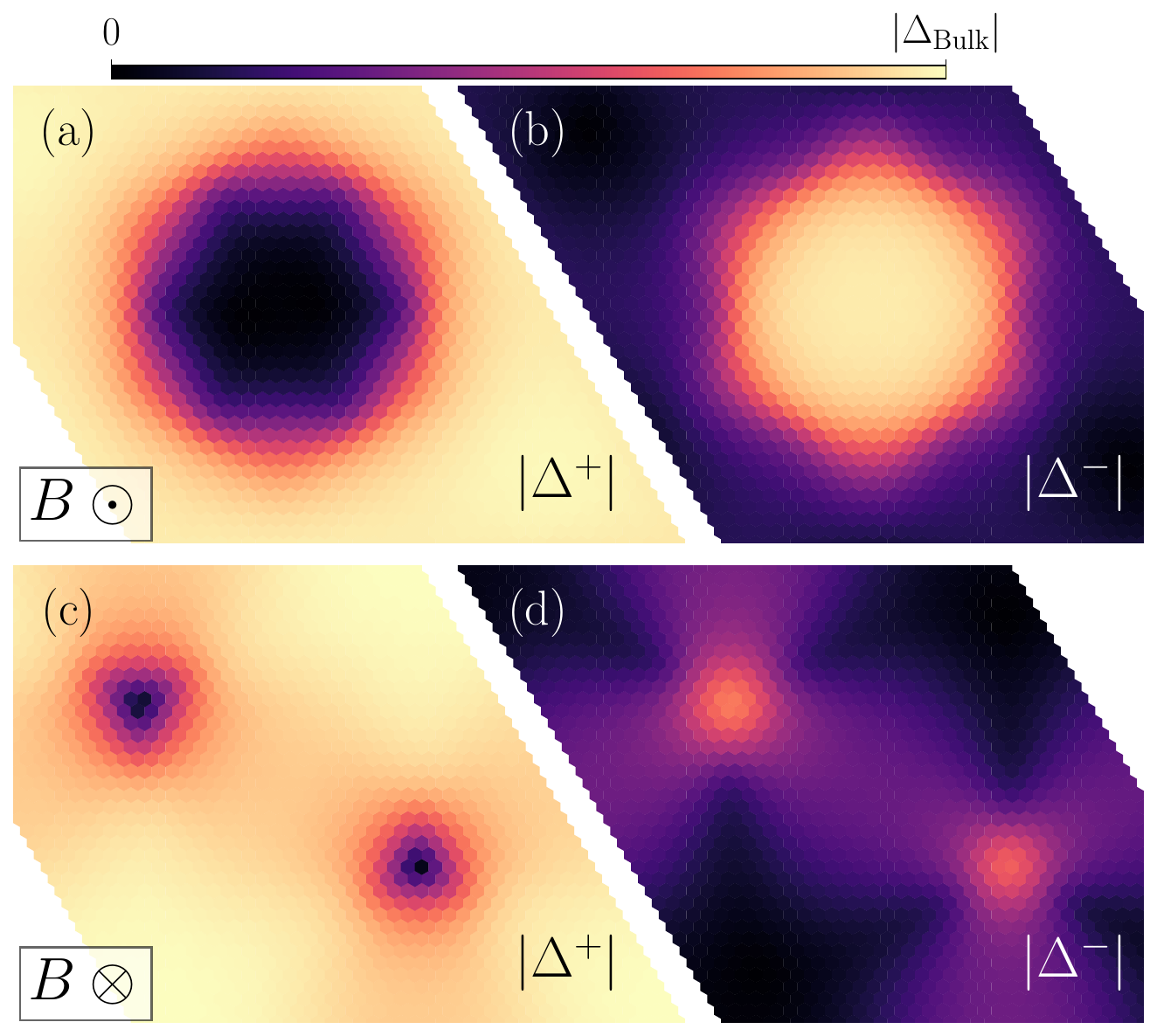}
    \caption{Magnitude of the two chiral order parameters $\Delta^{\pm}$ for each unit cell for the vortex solutions found at the upper van Hove singularities. The first (second) row refers to the solution with the magnetic field in the $\odot$ ($\otimes$) direction. The columns (left to right) correspond to the two chiralities $|\Delta^{+}|$ and $|\Delta^{-}|$.
    In the region enclosed by the fractional vortices, the bulk chirality $\Delta^+$ is suppressed and the opposite chirality $\Delta^-$ is induced, forming a closed domain wall.
    }
    \label{fig:UvH_chiral_OP}
\end{figure}

Next we investigate the spatial structure of the superconducting field $\Delta$. 
FIG.~\ref{fig:UvH_chiral_OP} shows the magnitude of the chiral order parameters, defined by Eq.~\eqref{Eq:chiral_UvH_state}, for both field directions. 
In both cases, the bulk chirality $\Delta^{+}$ is suppressed inside the region enclosed by the fractional vortices.
In the same region, the opposite chirality $\Delta^{-}$ is induced.
The boundary of this region forms a closed domain wall, on top of which the fractional vortices are located.
There is a free energy cost associated with introducing such a domain wall (which scales linearly with the domain wall boundary length).
This energy cost is compensated by the repulsive interaction of fractional vortices on domain walls, which makes these states energetically stable.
The fractional vortices together with the closed domain wall form a collective structure referred to as a ``coreless vortex'' (since the total superconducting field never vanishes), which can carry more than one superconducting flux quantum (as in the case for the $\odot$ field direction).
The formation of such collective structures has previously been studied and found to be energetically stable in TRSB superconductors~\cite{Matsunaga, Ichioka2005,Chung_2009,
babaev_garaud_2012_prl,
CP2_garaud_carlstrom_babaev_2013, 
babaev_garaud_2015,holmvall_2023_short, CP2_benfenati_2023}.

\subsubsection{Topological charge density}

Next we discuss and study the topological nature of the vortex states presented above.
In single-component superconductors, described by a scalar complex field $\psi (\textbf{r})$, the topological invariant is the winding number of $\psi$.
For multicomponent systems, let us denote the order parameter as $\Vec{\Psi} (\textbf{r}) = (\psi_1, \psi_2, \ldots , \psi_N)$.
This applies to our model of superconductivity on the kagome lattice, where each unit cell contains 9 singlet bonds (three from each sublattice).
Similar to single-component systems, multicomponent superconductors can form conventional Abrikosov vortices if the multicomponent order parameter can be written as $\Vec{\Psi}(\textbf{r}) = f(\textbf{r}) \Vec{\Psi}_0$, where $f$ is a complex-valued scalar field vanishing at some point $f(\textbf{r}_0)=0$. The topological invariant associated with this multicomponent Abrikosov vortex is the winding number of $f$.
However, the multicomponent nature allows for fundamentally different vortex states.
As shown in FIG.~\ref{fig:UvH_chiral_OP}, there exists no point in space where $\Vec{\Psi}$ vanishes, demonstrating that these cannot be thought of as conventional vortices.
This means that the topological invariant for these states cannot be attributed to a winding number of a single phase.
It is still possible to compute the winding number and locate the vortex core position of the individual components, however assigning these individual vortices with particular observables (such as the LDOS and currents) needs to be done carefully.
The reason for this is that both the values of the individual winding numbers and the vortex core locations are highly dependent on the choice of basis for $\Vec{\Psi}$.
This is demonstrated clearly for a $d+id$ superconductor in Refs.~\cite{holmvall_2023_short, holmvall_2023_long}, which explicitly demonstrate different winding numbers and fractional vortex locations depending on whether the chosen basis is $(\psi_{xy}, \psi_{x^2-y^2})$ or $(\psi_+, \psi_-)$, where $\psi_\pm = (\psi_{xy} \pm i \psi_{x^2 - y^2}) / \sqrt{2}$.
The same applies to the present case. A formal and systematic approach to identify the topological nature of the states is not to rely on phase windings of the individual components, but rather study the non-trivial topological structure embedded in the relative degrees of freedom of the order parameter.
Importantly, this topological structure is independent of the basis choice and is captured in a $\mathbb{C}P^{N-1}$ topological invariant (where $\mathbb{C}P^{N-1}$ denotes the complex projective space of complex vectors in $\mathbb{C}^N$).
These coreless vortex states are therefore sometimes also referred to as $\mathbb{C}P^{N-1}$ skyrmions.
Previous work on such skyrmions focused on nonlinear sigma-models 
\cite{Golo_1978, Dadda_1978, Akagi_2021_sigma_model_DMI, Akagi_2021_fractional_skyrme},
but has now been discussed in condensed matter settings, mainly in magnetism 
\cite{Kovrizhin_2013_skyrmion, Lian_2017_skyrmion_graphene, Zhang_2023_CP2_magnet, Amari_2022_CP2_DMI}, but also in superconductivity
\cite{CP2_garaud_carlstrom_babaev_2013, CP2_benfenati_2023} and superfluidity 
\cite{Eto_2013_bose_einstein}.

\begin{figure}
    \centering
    \includegraphics[width=1\linewidth, trim= 2.5cm 0cm 2.5cm 0cm,clip ]{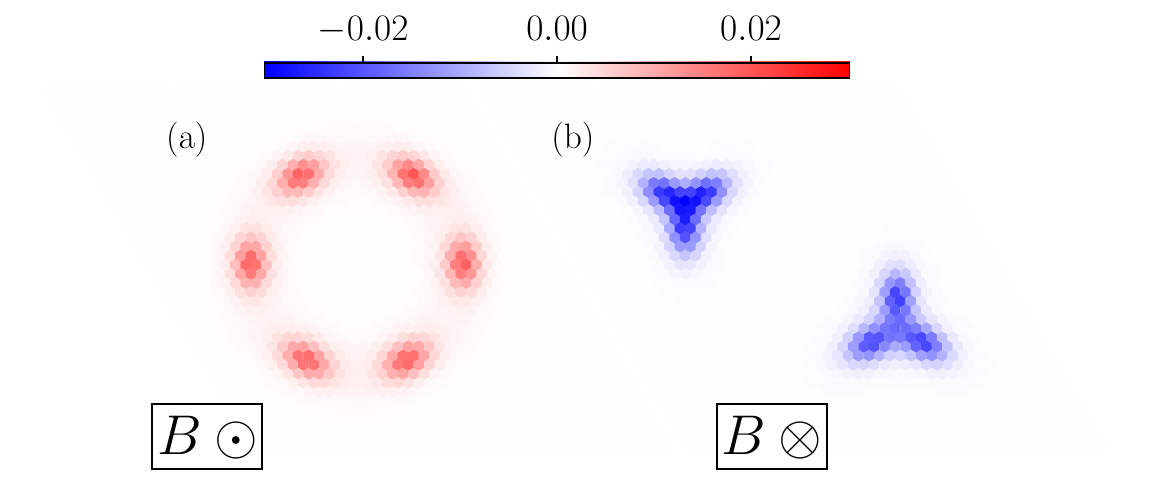} 
    \caption{The topological charge density for the two vortex solutions at the UvHS with the out-of-plane $\odot$ (a) and into-plane $\otimes$ field directions (b).
    The total topological charge is $\mathcal{Q} = +2$ and $\mathcal{Q} =-2$, respectively.
    }
    \label{fig:Skyr_UvH}
\end{figure}

We will now explain how to compute this topological invariant and its associated density, which can be used as a basis-independent indicator where in space the fractional vortices are located.
Since we are not interested in the total density of $\Vec{\Psi}$, we can normalize and define $\Vec{Z}(\textbf{r}) = \Vec{\Psi} (\textbf{r}) / |\Vec{\Psi} (\textbf{r})|$. 
Note that $\Vec{\Psi}$ and consequently $\Vec{Z}$ is defined on each unit cell.
The topological charge density associated with a face consisting of $n$ unit cells with locations $\textbf{r}_1, \textbf{r}_2, \ldots , \textbf{r}_n$ is related to the phase of the Bargmann invariant \cite{Bargmann_1964_wigner, Berg_topological_number_lattice, CP2_benfenati_2023}
\begin{equation}\label{Eq:skyr_density}
    \rho_{12\ldots n} = \frac{1}{2\pi} \arg\left( \prod_{i=1}^n \Vec{Z}_{i}^\dagger \Vec{Z}_{i+1} \right),
\end{equation}
where $\Vec{Z}_i = \Vec{Z}(\textbf{r}_i)$ and $\Vec{Z}_{n+1} = \Vec{Z}_1$. 
This local topological charge density can be viewed as the discrete version of the integral of the Berry curvature over the face $(1,2,\ldots, n)$ \cite{Berry_1984, Samuel_1988_geometric, Fukui_2005_chern}.
It is clear that this density is invariant under any change of basis $\Vec{\Psi} \mapsto U \Vec{\Psi}$, where $U$ denotes a spatially independent unitary matrix. The total topological invariant is the sum of the local charge densities, where we sum over the faces which tile the plane.
For the two different field directions, we find a topological invariant of $\mathcal{Q} = \pm 2$. 
That is, the topological invariant is precisely connected to the magnetic flux though the system $\Phi / \Phi_0 = \pm 2$, consistent with the expected flux quantization from continuum theory \cite{CP2_garaud_carlstrom_babaev_2013}.
The local topological charge density is shown in FIG.~\ref{fig:Skyr_UvH} 
\footnote{Here we compute a density for each unit cell, where the density is associated with the face consisting of its six nearest neighbor unit cells. 
Since these faces combined tile the plane three times, we need to compensate with a factor $1/3$ in the densities. 
This is merely a technically, but allows us to visualize the density at a higher resolution.}.
The peaks in the topological charge density align well with the locations of enhanced LDOS and currents, demonstrating its connection to these observable quantities, and consistent with the fractional nature of the vortices.
 
\begin{figure}
    \centering
    \includegraphics[width=1\linewidth, trim= 0cm 0cm 0cm 0cm,clip ]{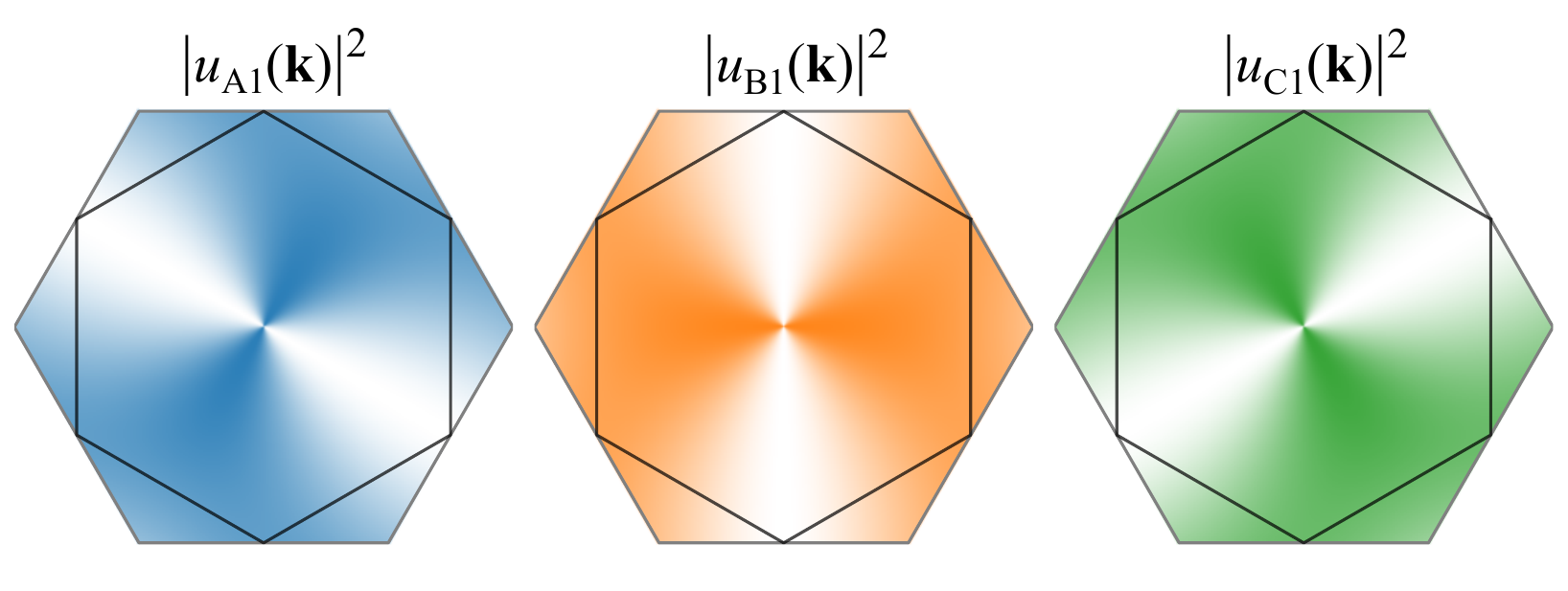} 
    \caption{Weights of the Bloch states in the lower band ($n=1$) on the A, B and C sublattices in the Brillouin zone. The black line segments form the Fermi surface for $\mu=-2$, which is identical in shape to that at $\mu=0$ in the upper band, see FIG.~\ref{fig:Basic_kagome_plots}. 
    }
    \label{fig:Sublattice_weights_LvH}
\end{figure}

\begin{figure*}
    \centering
    \includegraphics[width = 1 \linewidth]{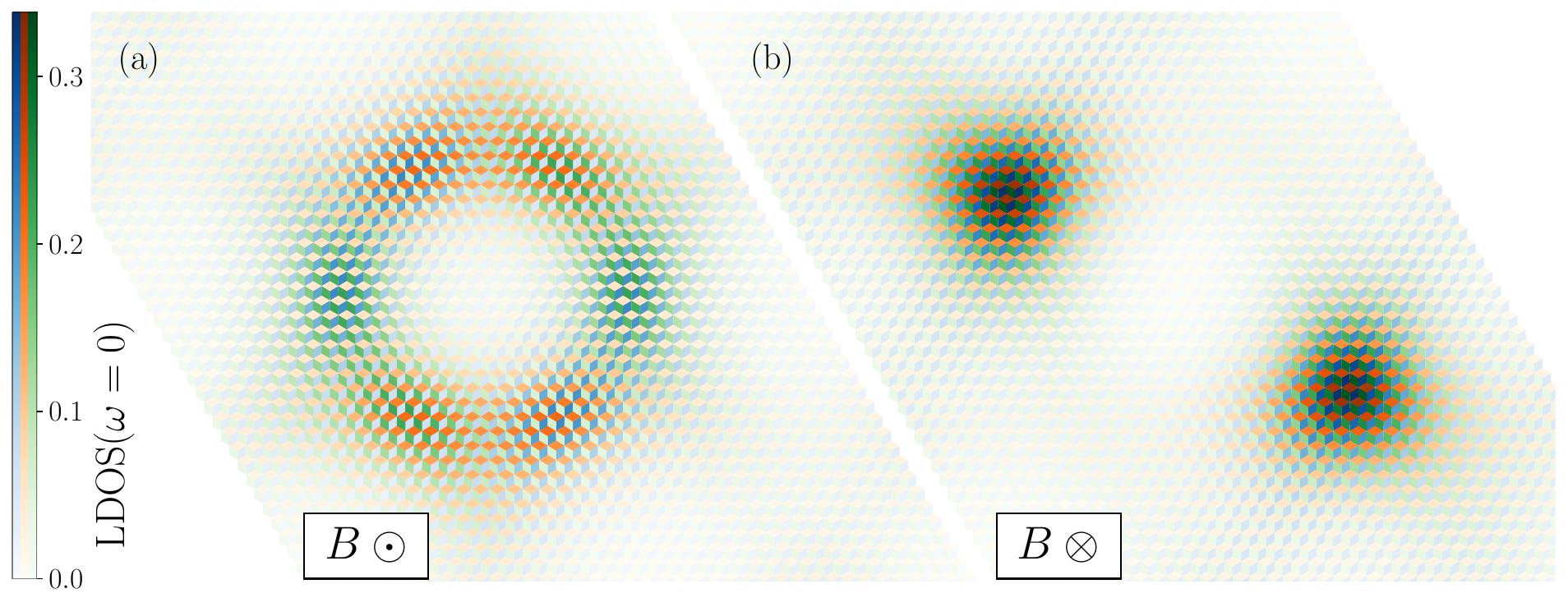}
    \caption{Local density of states at the Fermi level for the two field directions at the LvHS (corresponding LDOS at the UvHS shown in FIG.~\ref{fig:LDOS_UvH}).
    For both field directions, the bulk chirality is $\Delta^+$ and two superconducting flux quanta $\pm2\Phi_0$ penetrate the system. 
    For the out-of-plane ($\odot$) field direction, 6 regions of enhanced LDOS are observed, with an inverted sublattice character compared to the UvHS in FIG.~\ref{fig:LDOS_UvH}. 
    For the into-plane ($\otimes$) field direction, only two regions of enhanced LDOS are observed with an equal amount from all sublattices.
    }
    \label{fig:LDOS_LvH}
\end{figure*}

\subsection{Vortex states at the lower van Hove singularity}\label{sec:vortex_LvH}
From the vortex states found near the UvHS, we observed features characteristic to the sublattice interference of the upper band.
To further investigate the role sublattice interference plays in vortex formation, we carry out a similar analysis near the LvHS.
The LvHS has precisely the same Fermi surface and density of states as the UvHS and also favors chiral $d$-wave superconducting order.
The only difference in the normal electronic state is the sublattice interference. Any difference between the vortex states at the LvHS and the UvHS will be directly related their distinct microscopic sublattice structure of the Bloch states.

Where the weight of the Bloch state at the UvHS is maximal (minimal) for a given sublattice, the weight is minimal (maximal) on the same sublattice at the LvHS.
At the $M$-points, where the Bloch states of the UvHS are completely localized on one sublattice, the Bloch states of the LvHS have no weight on that same sublattice.
This is seen in FIG.~\ref{fig:Sublattice_weights_LvH} where the weight of the Bloch states of the $n=1$ band is shown in the Brillouin zone.
Compared to the UvHS, this means that each sublattice degree of freedom cannot be as easily decoupled at the LvHS.

\begin{figure}[b]
    \centering
    \includegraphics[width=1\linewidth, trim= 2.5cm 0cm 2.5cm 0cm,clip ]{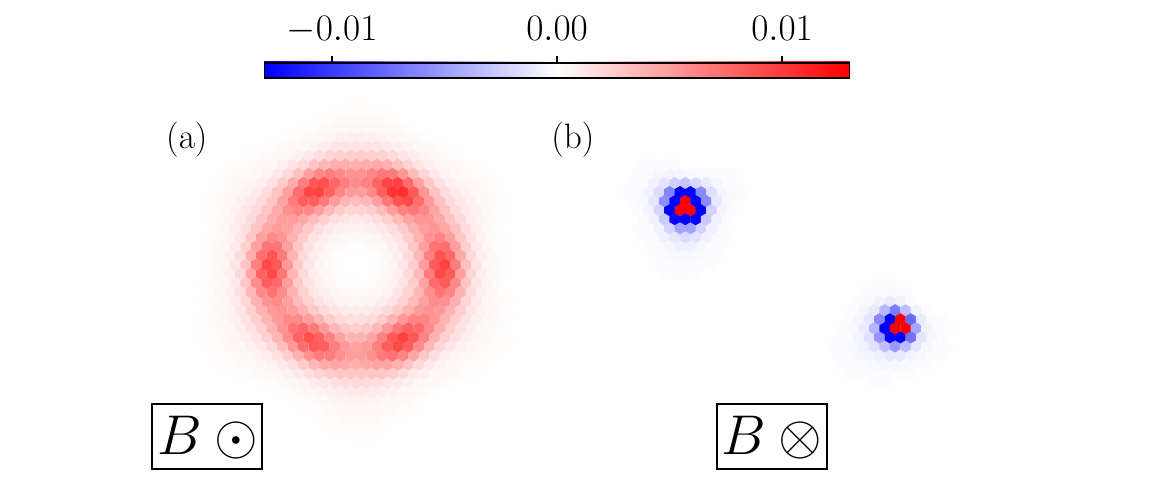} 
    \caption{
    The topological charge density for the two vortex solutions at the LvHS with the magnetic field out-of-plane $\odot$ in (a) and into-plane $\otimes$ in (b). The total topological charge is $\mathcal{Q} = +2$ in (a), indicating the presence of fractional vortices.
    In (b) the total topological charge is $\mathcal{Q}=0$, suggesting that this state consists of two conventional Abrikosov vortices.
    }
    \label{fig:Skyr_LvH}
\end{figure}

Performing similar self-consistent calculations at the LvHS, two unique solutions were obtained, one for each field direction. The LDOS for the two field directions are shown in FIG.~\ref{fig:LDOS_LvH}.
For an out-of-plane field we see a hexagonal LDOS feature, similar to the LDOS at the UvHS for the same field direction. 
However, a key difference between the UvHS and LvHS can be seen if we look at the sublattice character of the six LDOS peaks.
At the UvHS, the LDOS peaks were confined to a single sublattice, while at the LvHS, the peaks is in two of the sublattices.
For example, where the LDOS peak at the UvHS was only in sublattice A, there is a peak at the LvHS in B and C (i.e. not A).
This inverted sublattice character is seemingly inherited from the particular sublattice interference of the Bloch states at the van Hove singularities.

A closer inspection of the LDOS shows that in addition to the six distinct peaks (which form the hexagon), there is still  LDOS weight on the edges of the hexagon, which connects the dominant peaks.
These residual LDOS features raise questions regarding the interpretation of the six dominant peaks as separable fractional vortices. The peaks in the topological charge density, shown in FIG.~\ref{fig:Skyr_LvH}(a), align with the six regions of enhanced LDOS. However there is residual topological charge density in-between the peaks. This makes it less clear, compared to the UvHS, to interpret this state as consisting of six individual fractional vortices, each carrying a magnetic flux $\Phi_0 / 3$. In order to further clarify the nature of this vortex state, it would be useful to access even larger system sizes.

 The LDOS signal at the LvHS for the case opposite field direction $\otimes$ is shown in FIG.~\ref{fig:LDOS_LvH}(b). 
 It is clear that the maxima in the LDOS for all three sublattices overlap, indicating that these are two Abrikosov vortices. 
 The corresponding topological charge density is shown in FIG.~\ref{fig:Skyr_LvH}(b), which takes both positive and negative values, but integrates to zero \footnote{We cannot completely discard the possibility of there existing some stable unconventional vortex state since our method relies on initial guess.
 It could be the case that a stable unconventional vortex state with a particularly chosen initial guess.}. 
 This illustrates that both the sublattice interference (UvHS compared to LvHS) and the magnetic field direction ($\odot$ compared to $\otimes$) play a key role in the formation of the preferred vortex state.

\section{Discussion and Conclusions}\label{sec:conclusions}

In this work we investigated the vortex structure of a chiral $d$-wave superconducting state on the kagome lattice.
We used a fully self-consistent BdG formalism near electron fillings where the upper and lower van Hove singularities are close to the Fermi level, where the chiral TRSB $d$-wave state is favored.
Depending on the direction of the external magnetic field and which van-Hove singularity the Fermi level is close to, we found vortex states which are fundamentally different from the conventional Abrikosov vortex lattice.
Instead we observed the formation of coreless vortices, which consist of multiple fractional vortices each carrying one third of the superconducting flux quantum $\Phi_0 = h/2e$.
On one hand, the appearance of coreless and fractional vortices is not unexpected and have been discussed previously in multicomponent superconductors with spontaneous TRSB~\cite{Sauls01021994, tokuyasu_1990, Matsunaga, Ichioka2005,Chung_2009, babaev_garaud_2012_prl, babaev_garaud_2015, Haakansson2015,HasselbachUPt3,solitons_garaud_carlstrom_babaev_2011,
CP2_garaud_carlstrom_babaev_2013,
Garaud2014,
CP2_benfenati_2023,
holmvall_2023_short, holmvall_2023_long}.
On the other hand, finding $\Phi_0/3$ fractional vortices is exotic and deviates from the standard spontaneous symmetry-breaking Landau paradigm.
From a symmetry perspective, the two-fold degeneracy of the homogeneous ground state arises from the dimensionality of the irreducible representation $E_{2g}$ of our point group.
If an effective two-component Ginzburg-Landau (GL) theory is used (where the ground state could be written as $\Vec{\Psi}_\pm = (1, \pm i)$ in some basis), it would predict one-half flux quantum fractional vortices.
The one-third fractional vortices we find are self-consistent solutions to the mean-field equations of the BdG Hamiltonian, and demonstrate that key structure may get lost when going from the fully microscopic picture to the effective two-component GL field theory. 
What determines the fractionalization is not simply the dimensionality of the irrep - the microscopic details also can play a crucial role.
For the kagome lattice, this means the sublattice degrees of freedom and their interplay with the microscopic pairing mechanism.

Fractional vortices have been recently studied using self-consistent microscopic models in an $s+is$ superconductor \cite{CP2_benfenati_2023, iguchi_2023, Timoshuk_2025}, where TRSB superconductivity arises from interband repulsion in a multiband superconductor \cite{Stanev2010}.
The kagome lattice features three bands, but close to the van Hove singularities effectively only one band is present.
At the UvHS, we found fractional vortices belonging to individual sublattice degrees of freedom  (A, B or C). 
The active band at the UvHS exhibits sublattice interference of the Bloch states, such that the Fermi surface is extremely sublattice polarized. 
The Fermi surface can effectively be partitioned into three ``patches'', each associated with their own sublattice degree of freedom \cite{Scammell2023, Wu_2023_PDW_kagome}. 
This allows us to think of these fractional vortices as belonging to one of the three Fermi surface patches, establishing some connection to the multiband fractional vortices.
At the LvHS, the vortex state has an inverted sublattice character which is inherited from the inverted sublattice interference of the lower band compared to the upper.
At both van-Hove singularities, precisely the same symmetry is broken by the superconducting state and the shape of the Fermi surfaces is identical. 
The difference in the vortex structure between the UvHS and LvHS demonstrate clearly the importance of the microscopic character of the sublattice interference.

We stress that although we have focussed on vortex phases for two chemical potentials ($\mu = -0.02t$ for UvHS and $\mu=-1.98t$ for LvHS), we have additionally investigated the vortex states range of other chemical potentials $\mu \in [-0.2t,0.3t]$. We observed qualitatively similar solutions, which shows that the chemical potential does not need to be fine-tuned precisely to the van-Hove fillings in order to stabilize fractional vortices.

We briefly comment on isolated fractional vortices compared to coreless vortices.
The method of magnetic unit cells forces the magnetic flux through the magnetic unit cell to be $\Phi = 2n \Phi_0$, where $n$ is an integer. 
We can therefore, with this method, not find isolated fractional vortices, which only carry a fraction of the superconducting flux quantum $\Phi_0$. 
Instead we identify the fractional vortices as the building blocks for the coreless vortices, which carry integer-multiples of the superconducting flux quantum.
This does not mean that it is fundamentally impossible to find isolated fractional vortices. 
It is expected that individual fractional vortices should be stable on the boundary between pre-existing domains of different chirality.
Although of a fundamentally different nature than the fractional vortices we find on the kagome lattice, recent scanning SQUID measurements claim to resolve individual fractional vortices in ${\rm{Ba}}_{1-x}{\rm{K}}_x{\rm{Fe}}_2{\rm{As}}_2$ \cite{iguchi_2023} and $\rm{UPt}_3$ \cite{garcia_2025_UPt3_fractional}.

Finally, we point out that it is well-known that unconventional superconductors may feature unusual vortex structures. This has been discussed both in terms of vortex-induced competing phases in correlated unconventional superconductors~\cite{Arovas,Andersen2000,Chen2002,Ghosal2002,Zhu2002,Takigawa2003,Udby2006,Schmid2010,Andersen2011vortex,Agterberg2015}, and in terms of the presence of fractional vortices in multicomponent condensates~\cite{Sauls01021994, tokuyasu_1990, Matsunaga, Ichioka2005,Chung_2009, babaev_garaud_2012_prl, babaev_garaud_2015, solitons_garaud_carlstrom_babaev_2011,
CP2_garaud_carlstrom_babaev_2013,
Garaud2014,
CP2_benfenati_2023,
holmvall_2023_short, holmvall_2023_long}. The theoretical study presented in this paper focuses on fractional vortices in chiral $d$-wave kagome superconductors, and was motivated in part by the recently discovered vanadium-based kagome superconductors~\cite{Ortiz2020CsV3Sb5,OrtizEA21,YinEA21}.
We find that the LDOS is quantitatively different for magnetic fields into the plane compared to fields out of the plane, a property that was used as evidence of TRSB chiral kagome superconductivity from STM measurements~\cite{Deng2024}. Moreover, our results show that the different field directions give qualitatively distinct LDOS pattern due to the different arrangement of fractional vortices. However, our predictions of sublattice polarized fractional vortices applies more generally to other possible realizations of chiral $d$-wave condensates on the kagome lattice. We hope the present findings will motivate new experimental scanning SQUID microscopy or STM studies of the possible existence of fractional vortices in kagome superconductors. This may require careful investigations of magnetic field and temperature dependencies in order to locate favorable conditions, overcoming pinning effects while allowing for domain wall formations between different chiral regions. The existence of fractional vortices would be strong evidence of a chiral multicomponent ground state condensate.

\begin{acknowledgments}
F.A.S.P and B.M.A. acknowledge support from the Independent Research Fund Denmark Grant No. 5241-00007B. M.B was supported by a research grant (VIL69220) from VILLUM FONDEN. A.K. acknowledges support by the Danish National Committee for Research Infrastructure (NUFI) through the ESS-Lighthouse Q-MAT. 
\end{acknowledgments}

\bibliography{references}
\end{document}